\documentclass[fleqn]{revtex4}

\usepackage{amsmath}

\begin{document}

\title{Leading order relativistic corrections to the dipole polarizability of the hydrogen molecular ions.}

\author{D.T.~Aznabayev}
\author{A.K.~Bekbaev}
\author{S.A.~Zhaugasheva}
\affiliation{Bogoliubov Laboratory of Theoretical Physics, Joint Institute
for Nuclear Research, Dubna 141980, Russia}
\affiliation{Kazakh National University after al--Farabi, al-Farabi 71 Ave.,050038, Almaty,  Kazakhstan}

\author{V.I.~Korobov}
\affiliation{Bogoliubov Laboratory of Theoretical Physics, Joint Institute
for Nuclear Research, Dubna 141980, Russia}

\begin{abstract}
The static dipole polarizability for the hydrogen molecular ions H$_2^+$, HD$^+$, and D$_2^+$ are calculated. These new data for polarizability takes into account the leading order relativistic corrections to the wave function of the three-body system resulted from the Breit-Pauli Hamiltonian of $m\alpha^4$ order. Our study covers a wide range of rotational ($L\!=\!0\!-\!5$) and vibrational ($v\!=\!0\!-\!10$) states, which are of practical interest for precision spectroscopy of the hydrogen molecular ions.
\end{abstract}

\maketitle

\section{Introduction}

It has been shown recently \cite{SchillerPRL14} that simple molecular ions have potentiality to be used as optical clocks with very good stability. An essential ingredient for high fidelity of such clocks is a good knowledge of the molecule reaction on the external fields appeared in the experimental setup. For the hydrogen molecular ions (HMI) as the simplest three-body system such data can be rigorously obtained from the \emph{ab initio} calculations with a very high precision. The nonrelativistic polarizability of ro-vibrational states with up to eight or sometimes even more significant digits is now available for a wide range of states \cite{Hilico01,Pilon12,China15,SchillerPRA14}. It is easy to show that relativistic corrections to polarizability enters at a relative order of $\mathcal{O}(\alpha^2)$ or contribute to about $10^{-4}$ of relative precision \cite{relpol01}.

Furthermore there are many experiments carried out at present with the aim to get better determination of the proton to electron mass ratio using the ro-vibrational transition spectroscopy \cite{Koelemeij07,Koelemeij16} and to study the hyperfine structure of the HMI states \cite{Shen12,Bressel12}. The present status of theory for the ro-vibrational transitions is the fractional uncertainty of $\>\sim\!7\times10^{-12}$ for the fundamental transitions in HMI \cite{KorobovPRA14} and for the hyperfine structure precision achieved is at 1 ppm level \cite{KorobovPRL15}.

In this work we carry out calculations of the leading order relativistic correction to the dipole polarizability of the hydrogen molecular ions H$_2^+$, HD$^+$, and D$_2^+$. We take into consideration a wide range of ro-vibrational states: $L=0\!-\!5$, $v=0\!-\!10$. This is the first systematic study of the relativistic effects to polarizability of the HMIs for a variety of states. The higher order ($\mathcal{O}(\alpha^3)$) radiative corrections in principle may be also taken into account, so far that has been done rigorously only for the helium ground state \cite{Pachucki00,Szalewicz04}. In our present work we do not touch this issue, thus our current predictions are limited to 6-7 significant digits as physically meaninful quantity.

We adopt atomic units ($m_e=\hbar=e=1$) throughout this work.

\section{Theory}

\subsection{Nonrelativistic dipole polarizability}

We start from the nonrelativistic Schr\"odinger equation:
\begin{equation}\label{eq:NR}
(H_0-E_0)\Psi_{0}=0,
\qquad 
H_{0} = 
   \frac{\mathbf{P}_1^2}{2M_1}+\frac{\mathbf{P}_2^2}{2M_2}
   -\frac{\mathbf{p}_e^2}{2m_e}
   +\frac{Z_1Z_2}{R}-\frac{Z_1}{r_{1}}-\frac{Z_2}{r_{2}},
\end{equation}
where $\mathbf{P}_i$ and $M_i$ are impulses and masses of nuclei (proton or deuteron), $R$ is the internuclear distance, $r_{1}$ and $r_{2}$ are the distances from nuclei 1 and 2 to the electron, respectively. $Z_1$ and $Z_2$ are charges of the nuclei, in what follows we assume $Z_1=Z_2=Z$. The nonrelativistic state $\Psi_{0}=|v\, L\rangle$ is the unperturbed state characterized by the vibrational and rotational quantum numbers $v,\, L$, and $E_{0}$ is the state energy.

The interaction with an external electric field $\boldsymbol{\mathcal{E}}$ in the electric dipole representation is taken in the form
\begin{equation}
V_p = -\boldsymbol{\mathcal{E}}\cdot{\bf d},
\qquad
{\bf d} = e\bigl[Z(\mathbf{R}_{1}+\mathbf{R}_{2})-\mathbf{r}\bigr]\,,
\end{equation}
where $\mathbf{d}$ is the electric dipole moment of the HMI, and ${\bf R}_{1,2}$ and $\mathbf{r}$ are the position vectors of the nuclei and of electron with respect to the center of mass of the ion.

The change of energy due to polarizability of molecular ions is expressed by
\begin{equation}
\begin{array}{@{}l}\displaystyle
E_p^{(2)} =
   \langle\Psi_0|V_p(E_0-H_0)^{-1}V_p|\Psi_0\rangle
\\[2mm]\displaystyle\hspace{8mm}
 = \mathcal{E}^i\mathcal{E}^j \langle\Psi_0|d^i(E_0-H_0)^{-1}d^j|\Psi_0\rangle
 = -\frac{1}{2}\alpha_d^{ij}\mathcal{E}^i\mathcal{E}^j,
\end{array}
\end{equation}
where $\alpha^{ij}$ is a tensor of polarizability of rank 2,
\begin{equation}
\alpha_d^{ij}=-2\langle\Psi_0|d^i(E_0-H_0)^{-1}d^j|\Psi_0\rangle.
\end{equation}

The static dipole polarizability tensor is then reduced \cite{Landau} to scalar, $\alpha_s$, and tensor, $\alpha_t$, terms, which may be expressed by the three contributions corresponding to the possible values of $L'$ for the angular momentum of the intermediate state: $L'\!=\!L\!\pm\!1$, or $L'\!=\!L$.
\begin{equation}
\begin{array}{@{}l}
\displaystyle a_+= \;\;\frac{2}{2L+1}
\sum_n\frac{\langle 0L\|\mathbf{d}\|n(L\!+\!1) \rangle
            \langle n(L\!+\!1)\|\mathbf{d}\|0L \rangle}{E_0-E_n},\\[4mm]
\displaystyle a_0=-\frac{2}{2L+1}
\sum_n\frac{\langle 0L\|\mathbf{d}\|nL \rangle
            \langle nL\|\mathbf{d}\|0L \rangle}{E_0-E_n},\\[4mm]
\displaystyle a_-= \;\;\frac{2}{2L+1}
\sum_n\frac{\langle 0L\|\mathbf{d}\|n(L\!-\!1) \rangle
            \langle n(L\!-\!1)\|\mathbf{d}\|0L \rangle}{E_0-E_n}.
\end{array}
\end{equation}
where $E_n$ is the energy of the intermediate state $|nL'\rangle$. The polarizability tensor operator on a subspace of fixed total orbital angular momentum $L$ may now be expressed:
\begin{equation}\label{alpha:op}
\alpha_d^{ij} = \alpha_s+\alpha_t
                \left[L^iL^j+L^jL^i-\frac{2}{3}\mathbf{L}^2\right],
\end{equation}
where
\begin{equation}\label{alpha2}
\begin{array}{@{}l}
\displaystyle
\alpha_s^{} = \frac{1}{3}\bigl[a_++a_0+a_-\bigr],
\\[4mm] \displaystyle
\alpha_t^{} = -\frac{a_+}{2(L+1)(2L+3)}
              +\frac{a_0}{2L(L+1)}-\frac{a_-}{2L(2L-1)}\>.
\end{array}
\end{equation}

The basic formalism in a shorthand vector notation may be rewritten as follows
\begin{equation}\label{pol_short}
\begin{array}{@{}l}\displaystyle
\Psi_1 = (E_0\!-\!H_0)^{-1}\mathbf{d}|\Psi_0\rangle
\\[3mm]\displaystyle
\alpha_d = \left\langle\Psi_0|\mathbf{d}|\Psi_1\right\rangle
         = \left\langle\Psi_0|\mathbf{d}(E_0\!-\!H_0)^{-1}\mathbf{d}|\Psi_0\right\rangle.
\end{array}
\end{equation}

The Tables for a complete set of the nonrelativistic results for H$_2^+$, HD$^+$ and D$_2^+$ ions within the range of the ro-vibrational states under consideration are presented in \cite{SchillerPRA14}.

\begin{table}
\begin{center}
\caption{Test of convergence of the numerical results. The groung state of H$_2^+$ molecular ion is chosen for consideration. $N$ is the number of basis functions for initial and intermediate states used in calculations. For convenience of comparison with other authors the values of nuclear masses have been taken: $M_p = 1836.152701\>m_e$ and $M_d = 3670.483014\> m_e$.}\label{conv}
\begin{tabular}{l@{\hspace{15mm}}l@{\hspace{10mm}}l@{\hspace{10mm}}l}
\hline\hline
$N$ & ~~~~~~~~~~~~~~~$E_{\rm NR}$ & ~~~~~~~~~~$\alpha_d$ & $(1/c)^2\alpha_B\times10^2$ \\
\hline
 2000 & $-0.59713\>90631\>23404\>0757$ & $3.16872\>58022\>7017$ & $-1.52753848$ \\
 3000 & $-0.59713\>90631\>23405\>0374$ & $3.16872\>58026\>7529$ & $-1.52753844$ \\
 4000 & $-0.59713\>90631\>23405\>0730$ & $3.16872\>58026\>7610$ & $-1.52753841$ \\
 5000 & $-0.59713\>90631\>23405\>0747$ & $3.16872\>58026\>7613$ & $-1.52753839$ \\
\hline\hline
\end{tabular}
\end{center}
\end{table}

\begin{table}
\begin{center}
\caption{Poralizability of $\mbox{H}_2^+$ molecular ion.} \label{H2+}
\hspace*{-5mm}\footnotesize
\begin{tabular}{r@{\hspace{3mm}}c@{\hspace{3mm}}cc@{\hspace{3mm}}cc@{\hspace{3mm}}cc@{\hspace{3mm}}cc@{\hspace{3mm}}cc}
\hline\hline
\vrule width0pt height 12pt
& $L=0$ & \multicolumn{2}{c}{$L=1$} & \multicolumn{2}{c}{$L=2$}
& \multicolumn{2}{c}{$L=3$} & \multicolumn{2}{c}{$L=4$} & \multicolumn{2}{c}{$L=5$}  \\
\cline{3-4}\cline{5-6}\cline{7-8}\cline{9-10}\cline{11-12}
$v$ & $\alpha_s$ & $\alpha_s$ & $\alpha_t$ & $\alpha_s$ & $\alpha_t$ & $\alpha_s$
    & $\alpha_t$ & $\alpha_s$ & $\alpha_t$ & $\alpha_s$ & $\alpha_t$ \\
\hline
\vrule width0pt height 11pt
 0 & 3.1685731 & 3.1781425 &$-$0.8033502 & 3.1973545 &$-$0.1931356 & 3.2262879 &$-$0.0914433 & 3.2650990 &$-$0.0544748 & 3.3139976 &$-$0.0367128 \\
 1 & 3.8973934 & 3.9099178 &$-$1.1441799 & 3.9350819 &$-$0.2750942 & 3.9730164 &$-$0.1302617 & 4.0239695 &$-$0.0776116 & 4.0882763 &$-$0.0523165 \\
 2 & 4.8213113 & 4.8378793 &$-$1.6000406 & 4.8711902 &$-$0.3847577 & 4.9214594 &$-$0.1822335 & 4.9890778 &$-$0.1086134 & 5.0745756 &$-$0.0732459 \\
 3 & 6.0091177 & 6.0313112 &$-$2.2129254 & 6.0759600 &$-$0.5322677 & 6.1434165 &$-$0.2521933 & 6.2342968 &$-$0.1503867 & 6.3494400 &$-$0.1014829 \\
 4 & 7.5602216 & 7.5903867 &$-$3.0434518 & 7.6511105 &$-$0.7322788 & 7.7429690 &$-$0.3471380 & 7.8669387 &$-$0.2071473 & 8.0243574 &$-$0.1399094 \\
 5 & 9.6215210 & 9.6632217 &$-$4.1811193 & 9.7472225 &$-$1.0064538 & 9.8744707 &$-$0.4774294 & 10.046534 &$-$0.2851531 & 10.265571 &$-$0.1928167 \\
 6 & 12.415730 & 12.474532 &$-$5.7615463 & 12.593067 &$-$1.3876642 & 12.772916 &$-$0.6588237 & 13.016644 &$-$0.3939470 & 13.327789 &$-$0.2667717 \\
 7 & 16.290723 & 16.375602 &$-$7.9965248 & 16.546853 &$-$1.9273285 & 16.807161 &$-$0.9160282 & 17.160827 &$-$0.5485430 & 17.613818 &$-$0.3721506 \\
 8 & 21.809221 & 21.935211 &$-$11.228720 & 22.189694 &$-$2.7088006 & 22.577348 &$-$1.2892136 & 23.105612 &$-$0.7734482 & 23.784904 &$-$0.5259745 \\
 9 & 29.920158 & 30.113640 &$-$16.036365 & 30.504982 &$-$3.8730665 & 31.102663 &$-$1.8465662 & 31.920118 &$-$1.1104628 & 32.976304 &$-$0.7574536 \\
10 & 42.306376 & 42.616282 &$-$23.446097 & 43.244214 &$-$5.6711694 & 44.206322 &$-$2.7100353 & 45.528226 &$-$1.6347899 & 47.246401 &$-$1.1195399 \\
\hline\hline
\end{tabular}
\end{center}
\end{table}

\subsection{Relativistic corrections to the dipole polarizability}

Relativistic correction $\alpha_B$ to the static dipole polarizability: 
\begin{equation}
\alpha_d^{} = \alpha_d^{(\rm nonrel)}+(1/c)^2\,\alpha^B,
\end{equation}
is expressed:
\begin{equation}
\begin{array}{@{}l}\displaystyle
\alpha_B = 2\left\langle\Psi_B|\mathbf{d}|\Psi_1\right\rangle
          + \left\langle\Psi_1\left|H_B\!-\!\langle H_B \rangle\right|\Psi_1\right\rangle
\\[2mm]\displaystyle\hspace{6mm}
 = 2\left\langle\Psi_0|H_BQ(E_0\!-\!H_0)^{-1}Q\mathbf{d}|(E_0\!-\!H_0)^{-1}\mathbf{d}\Psi_0\right\rangle
\\[2mm]\displaystyle\hspace{20mm}
   + \left\langle\Psi_0\left|\mathbf{d}(E_0\!-\!H_0)^{-1}
   \left(H_B\!-\!\langle H_B \rangle\right)
   (E_0\!-\!H_0)^{-1}\mathbf{d}\right|\Psi_0\right\rangle,
\end{array}
\end{equation}
here $H_B$ is the Hamiltonian of the Breit-Pauli interaction for the three-body system \cite{HB,BPrelcor}:
\begin{equation}
\begin{array}{@{}l}\displaystyle
H_B = 
      -\frac{\mathbf{p}_e^4}{8m_e^3}
      +\frac{4\pi }{8m_e^2}
         \left[Z_1\delta(\mathbf{r}_1)
              +Z_2\delta(\mathbf{r}_2)
         \right]
      -\frac{\mathbf{P}_1^4}{8M_1^3}-\frac{\mathbf{P}_2^4}{8M_2^3}
\\[3mm]\displaystyle\hspace{15mm}
      +\frac{Z_1}{2m_eM_1}
      \left(
         \frac{\mathbf{p}_e\mathbf{P}_1}{r_1}
         +\frac{\mathbf{r}_1(\mathbf{r}_1\mathbf{p}_e)\mathbf{P}_1}{r_1^3}
      \right)
      +\frac{Z_2}{2m_eM_2}
      \left(
         \frac{\mathbf{p}_e\mathbf{P}_2}{r_2}
         +\frac{\mathbf{r}_2(\mathbf{r}_2\mathbf{p}_e)\mathbf{P}_2}{r_2^3}
     \right)
\\[3mm]\displaystyle\hspace{30mm}
     -\frac{Z_1Z_2}{2M_1M_2}
     \left(
        \frac{\mathbf{P}_1\mathbf{P}_2}{R}
        +\frac{\mathbf{R}(\mathbf{R}\mathbf{P}_1)\mathbf{P}_2}{R^3}
     \right),
\end{array}
\end{equation}
and $\Psi_B$ is the relativistic correction to the nonrelativistic wave function $\Psi_0$:
\begin{equation}\label{eq:HB}
\Psi_B = Q(E_0\!-\!H)^{-1}QH_B|\Psi_0\rangle.
\end{equation}
Operator $Q$ in the above equations is the projection operator on the subspace orthogonal to $|\Psi_0\rangle$. It is worthy to note that in the atomic units $c = \alpha^{-1}$, where $\alpha$ is the fine structure constant. Expressions (\ref{pol_short}) and (\ref{eq:HB}) represent linear equations for the wave functions $\Psi_1$ and $\Psi_B$, respectively. And thus the final value for $\alpha_B$ may be obtain without full diagonalization of the Hamiltonian (\ref{eq:NR}) on the subspace of the variational basis set and futher summation over states and pseudostates, that makes numerical procedure more fast and stable.

It is easily seen that the relativistic correction to the polarizability energy shift is the third order contribution of the pertubation theory.  It is linear in the parameter $\alpha^2$, natural parameter of the relativistic expansion, and is quadratic in the electric field density $\boldsymbol{\mathcal{E}}$.

\begin{table}
\begin{center}
\caption{Poralizability of $\mbox{D}_2^+$ molecular ion.} \label{D2+}
\hspace*{-5mm}\footnotesize
\begin{tabular}{r@{\hspace{3mm}}c@{\hspace{3mm}}cc@{\hspace{3mm}}cc@{\hspace{3mm}}cc@{\hspace{3mm}}cc@{\hspace{3mm}}cc}
\hline\hline
\vrule width0pt height 12pt
& $L=0$ & \multicolumn{2}{c}{$L=1$} & \multicolumn{2}{c}{$L=2$}
& \multicolumn{2}{c}{$L=3$} & \multicolumn{2}{c}{$L=4$} & \multicolumn{2}{c}{$L=5$}  \\
\cline{3-4}\cline{5-6}\cline{7-8}\cline{9-10}\cline{11-12}
$v$ & $\alpha_s$ & $\alpha_s$ & $\alpha_t$ & $\alpha_s$ & $\alpha_t$ & $\alpha_s$
    & $\alpha_t$ & $\alpha_s$ & $\alpha_t$ & $\alpha_s$ & $\alpha_t$ \\
\hline
\vrule width0pt height 11pt
0   &3.0718385  &3.0764328  &$-$0.7579298  &3.0856549  &$-$0.1813369  &3.0995081   &$-$0.0852409  &3.1180309
& $-$0.0502995 &3.1412711  &$-$0.0335034 \\
1   &3.5528638  &3.5584089  &$-$0.9782491  &3.5695458  &$-$0.2340523  &3.5862817   &$-$0.1100231  &3.6086695
& $-$0.0649249 &3.6367756  &$-$0.0432467 \\
2   &4.1194070  &4.1261334  &$-$1.2485728  &4.1396485  &$-$0.2987404  &4.1599666   &$-$0.1404395  &4.1871595
& $-$0.0828801 &4.2213195  &$-$0.0552122 \\
3   &4.7910944  &4.7993001  &$-$1.5808434  &4.8157922  &$-$0.3782640  &4.8405962   &$-$0.1778400  &4.8738102
& $-$0.1049648  &4.9155614  &$-$0.0699351 \\
4   &5.5931121  &5.6031855  &$-$1.9903702  &5.6234356  &$-$0.4762945  &5.6539046   &$-$0.2239563  &5.694727
& $-$0.1322053  &5.7460784  &$-$0.0881031  \\
5   &6.5581008  &6.5705539  &$-$2.4969746  &6.5955906  &$-$0.5975866  &6.6332784   &$-$0.2810323  &6.6838036
& $-$0.1659331  &6.7474075  &$-$0.1106091  \\
6   &7.7288214  &7.7443358  &$-$3.1265993  &7.7755297  &$-$0.7483664  &7.8225084   &$-$0.3520083  &7.8855305
& $-$0.2078938  &7.9649300  &$-$0.1386246  \\
7   &9.1619614  &9.1814571  &$-$3.9136098  &9.2206579  &$-$0.9368846  &9.2797220   &$-$0.4407829  &9.3590190
& $-$0.2604042  &9.4590076  &$-$0.1737066  \\
8   &10.933664  &10.958399  &$-$4.9041347  &11.008137  &$-$1.1742218  &11.083110   &$-$0.5525959  &11.183859
& $-$0.3265811  &11.311016  &$-$0.2179523  \\
9   &13.147705  &13.179427  &$-$6.1610100  &13.243217  &$-$1.4754799  &13.339457   &$-$0.6945965  &13.468847
& $-$0.4106816  &13.632332  &$-$0.2742296  \\
10  &15.947847  &15.989024  &$-$7.7712522  &16.071842  &$-$1.8615859  &16.196869   &$-$0.8766962  &16.365114
& $-$0.5186156  &16.577941  &$-$0.3465270  \\
\hline\hline
\end{tabular}
\end{center}
\end{table}

\begin{table}
\begin{center}
\caption{Poralizability of $\mbox{HD}^+$ molecular ion.} \label{HD+}
\footnotesize
\begin{tabular}{r@{\hspace{3mm}}c@{\hspace{3mm}}cc@{\hspace{3mm}}cc@{\hspace{3mm}}cc@{\hspace{3mm}}cc@{\hspace{3mm}}cc}
\hline\hline
\vrule width0pt height 12pt
& $L=0$ & \multicolumn{2}{c}{$L=1$} & \multicolumn{2}{c}{$L=2$}
& \multicolumn{2}{c}{$L=3$} & \multicolumn{2}{c}{$L=4$} & \multicolumn{2}{c}{$L=5$}  \\
\cline{3-4}\cline{5-6}\cline{7-8}\cline{9-10}\cline{11-12}
$v$ & $\alpha_s$ & $\alpha_s$ & $\alpha_t$ & $\alpha_s$ & $\alpha_t$ & $\alpha_s$
    & $\alpha_t$ & $\alpha_s$ & $\alpha_t$ & $\alpha_s$ & $\alpha_t$ \\
\hline
\vrule width0pt height 11pt
0 &395.27754  &3.9899486 &175.46989  &4.0093758 &13.826954  &4.0386030 &3.1905097  &4.0777634 &1.1013280  &4.1270299 &0.4731528 \\
1 &462.62017  &4.7029195 &205.18613  &4.7267328 &16.142249  &4.7625805 &3.7153046  &4.8106388 &1.2779012  &4.8711642 &0.5463772 \\
2 &540.64963  &5.5690125 &239.56394  &5.5984707 &18.814812  &5.6429097 &4.3189073  &5.7025090 &1.4799073  &5.7776380 &0.6295018 \\
3 &631.36149  &6.6325616 &279.45735  &6.6693071 &21.908555  &6.7251653 &5.0148197  &6.7999232 &1.7113981  &6.8942638 &0.7239099 \\
4 &737.27142  &7.9543692 &325.93821  &8.0010358 &25.503117  &8.0716799 &5.8197299  &8.1666401 &1.9772885  &8.2866362 &0.8312163 \\
5 &861.59725  &9.6180552 &380.37171  &9.6782520 &29.699534  &9.7691277 &6.7545125  &9.8914502 &2.2835914  &10.046247 &0.9533009 \\
6 &1008.5213  &11.742867 &444.52180  &11.821446 &34.627354  &11.940196 &7.8456173  &12.100244 &2.6377250  &12.303109 &1.0923369 \\
7 &1183.5771  &14.499937 &520.70922  &14.604304 &40.455664  &14.762192 &9.1270207  &14.975296 &3.0489087  &15.245914 &1.2507930 \\
8 &1394.2333  &18.141977 &612.04499  &18.283320 &47.409088  &18.497399 &10.643024  &18.786825 &3.5286700  &19.155146 &1.4313721 \\
9 &1650.8015  &23.051744 &722.79107  &23.247524 &55.792076  &23.544383 &12.452318  &23.946515 &4.0914633  &24.459532 &1.6367943 \\
10&1967.8945  &29.827372 &858.93232  &30.105532 &66.026733  &30.528154 &14.633990  &31.101840 &4.7553467  &31.835850 &1.8692259 \\
\hline\hline
\end{tabular}
\end{center}
\end{table}

\begin{table}
\begin{center}
\caption{Nonrelativistic static dipole polarizability. Comparison with previous calculations. For convenience of comparison the values of nuclear masses have been taken: $M_p = 1836.152701 m_e$ and $M_d = 3670.483014 m_e$.}
\begin{tabular}{l@{\hspace{20mm}}l@{\hspace{15mm}}l@{\hspace{15mm}}l}
\hline\hline
 & ~~~~~~~~H$_2^+$ & ~~~~~~~~D$_2^+$ & ~~~~~~~~HD$^+$ \\
\hline
Hilico {\it et al.} \cite{Hilico01} 
 & $3.16872\>5803$ & $3.07198\>8696$ & $395.30632\>88$ \\
Olivares Pil\'on, Baye \cite{Pilon12}
 & $3.16872\>58026\>5$ & & \\
Li-Yan Tang {\it et al.} \cite{TangPRA14}
 & $3.16872\>58026\>76(1)$ & $3.07198\>86956\>6(7)$ & $395.30632\>87972(1)^a$ \\
this work
 & $3.16872\>58026\>7613(1)$ & $3.07198\>86956\>7511(2)$ & $395.30632\>8797231(3)$ \\
\hline\hline
\end{tabular}
\end{center}
\flushleft{$^a$Zong-Chao Yan {\it et al.} \cite{Yan03}}
\end{table}

\section{Results}

In our calculations we use a variational method based on exponential expansion with randomly chosen exponents, which has been described in details in a variety of our previous works \cite{SchillerPRA14,BPrelcor} and we omit here an explicit formulation of the method.

First we study convergence of our numerical results. For demonstration we take the ground $(L\!=\!0,v\!=\!0)$ state of the H$_2^+$ molecular ion. As is seen from the Table \ref{conv}, the relativistic contribution $\alpha_B$ may be determined with at least eight significant digits while the nonrelativistic polarizability obtained is more precise than the best known in the literature \cite{TangPRA14}.

A complete set of data of our numerical calculations is collected in Tables \ref{H2+}--\ref{HD+}, the polarizabilities for H$_2^+$, D$_2^+$ and HD$^+$ molecular ions, respectively, and is the main result of the present work. In this case the values for masses of nuclei have been taken from the latest published adjustment \cite{CODATA10} of the CODATA group: $M_p = 1836.15267245\;m_e$ and $M_d = 3670.4829652\;m_e$. They are also in accordance with our previous nonrelativistic calculations \cite{SchillerPRA14}. To avoid numerical errors we have used the sextuple precision arithmetics (48 decimal digits).

In the last Table we compare our results with the previous ones. Due to absence of the data for the relativistic polarizability we include into out Table only the nonrelativistic values for polarizability. In all the cases our result demonstrate perfect agreement with previous calculations.

In conclusion we want to state that the new data for the polarizabilities of the hydrogen molecular ions have been obtained, which is significant up to 6-7 digits as physically observable quantities, while the achieved numerical precision is of eight or even more digits. These data may be used to increase precision of the physically meaningful values by including higher order QED corrections \cite{Pachucki00}. We want to note that this is the first systematic calculation, which includes the leading order relativistic corrections.

\section*{Acknowledgements}

The work has been carried out under financial support of the Program of the Ministry of Education and Science of the Republic of Kazakhstan 0263/PTF, which is gratefully acknowledged. V.I.K.\ also acknowledges support of the Russian Foundation for Basic Research under Grant No.~15-02-01906-a.

\end{document}